\documentclass[useAMS,usenatbib]{mn2e}

\usepackage[dvips]{graphicx}
\usepackage{psfrag}
\usepackage{amsmath}
\input{epsf}

\title[Synchronization mechanism of sharp edges in rings of Saturn]
{Synchronization mechanism of sharp edges in rings of Saturn}
\author[D. L. Shepelyansky, A. S. Pikovsky, J. Schmidt and F. Spahn]{
D. L. Shepelyansky$^{1,2,3}$\thanks{E-mail:
dima(at)irsamc.ups-tlse.fr (DLS); pikovsky(at)uni-potsdam.de (ASP); jschmidt(at)agnld.uni-potsdam.de (JS);
frank(at)agnld.uni-potsdam.de (FS)}, A. S. Pikovsky$^{3}$,
J. Schmidt$^{3}$ and  F. Spahn$^{3}$
\\
$^{1}$ Universit\'e de Toulouse, UPS, Laboratoire de Physique Th\'eorique
(IRSAMC), F-31062 Toulouse, France\\
$^{2}$ CNRS, LPT (IRSAMC), F-31062 Toulouse, France\\
$^{3}$ Department of Physics and Astronomy,
Universit\"at Potsdam,
Karl-Liebknecht-Str 24/25, Bld. 28
D-14476 Potsdam-Golm, Germany}
\begin{document}

\date{Accepted 200? XXXXXX YY. Received 2008 November XX; in original form 2008 November XX}

\pagerange{\pageref{firstpage}--\pageref{lastpage}} \pubyear{2009}

\maketitle

\label{firstpage}

\begin{abstract}
We propose a new mechanism which explains the existence of enormously sharp
edges in the rings of Saturn. This mechanism is based on the synchronization
phenomenon due to which the epicycle rotational phases of particles in the
ring, under certain conditions, become synchronized with the phase of 
external satellite, e.g. with the phase of Mimas in the case of the outer B ring edge.
This synchronization eliminates collisions between particles and suppress the
diffusion induced by collisions by orders of magnitude. The minimum of the
diffusion is reached at the center of the synchronization regime
corresponding to the ratio 2:1 between the orbital frequency at the edge of B
ring and the orbital frequency of Mimas. The synchronization theory gives the
sharpness of the edge in few tens of meters that is in agreement with available
observations.  
\end{abstract}


\section{Introduction}

Together with very small thickness, extreme sharp edges in rings of Saturn 
are one of the most outstanding features of planetary rings 
(see {\it e.g.} \citet{fridman,esposito1,esposito2,bgt82}). Indeed,
for example the outer B ring edge has a density drop by an order 
of magnitude on a distance $r_e \sim 10 m$ 
that is enormously sharp
compared to the edge distance to Saturn ($117580 km$),
the B ring width ($25580km$) and the width of Cassini division
($4620km$). This is especially surprising since the life time of the
rings is enormously large being of about $10^{12}$ orbital periods  \citep{fridman}
and the particles inside the B ring are quite dense 
(e.g. there are particles of size $10m$ down to $1cm$ and smaller
with a distance between them of about a few meters and less) \citep{fridman,esposito1,esposito2,spahn}.
Due to this about $10 -100$ collisions between particles per orbit should wash out all sharp density contrasts in only a few orbital periods.

In many cases, and perhaps always, these sharp edges are associated with the gravitational perturbation by a moon inside or outside the rings. 
The abruptness of the transition from nearly opaque to practically transparent 
regions was constrained by the Voyager Photopolarimeter data \citep{lane} 
to be smaller than about 100 meters. Cassini UVIS occultation profiles show edges sharper 
than tens of meters, and in fact the assumption of a knife-edge sharpness is used to 
constrain the local ring thickness near the edge to be on the order of 5 meters 
only (\citep{colwell08,colwell09}; ring thickness and sharpness of the edge should be of the same order).

After the pioneering work of \citet{wisdom}
extensive numerical simulations of particle dynamics
have been performed by different groups
(see e.g. \citet{salo1,stewart,salo2,porco,schmidt}) that allowed
to establish a number of interesting properties of ring dynamics.
However, the problem of sharp ring edges still remains
a mystery. Its solution requires extensive large scale numerical simulations
with particles of different scales, it also may require to go beyond local box simulations 
invented by \citet{wisdom}. In view of these difficulties
it seems to be useful to explore certain simplified models that bring
 to  surface new qualitative physical effects 
which can be analyzed more directly due to model simplicity.
Here, we introduce such a simplified model of ring dynamics for outer B ring edge
called the SYNC model in the following. Numerical investigations of this model
show a striking phenomenon of synchronization of the epicycle motions of
particles in  the ring induced, under certain conditions,  
by a periodical gravitational force of Mimas. In a general context, the synchronization phenomenon,
which has abundant manifestations in science, nature, engineering and social life
\citep{pikovsky,strogatz}, can be roughly described as an adjustment of frequencies and phases of oscillators due to interaction and/or forcing. In the present context,
the phases of the epicycle rotation of various particles
become synchronized with the phase of periodic gravitational force of Mimas,
and as a result 
the collisions between particles become suppressed by orders of magnitude
so that the 
diffusion in the ring also becomes suppressed by orders of magnitude.
This leads to a maintaince of sharp edges in a certain frequency range of ratios of
epicycle frequency $\Omega$ to the Mimas frequency $\omega$.
We note that the data of the Cassini mission show that
the outer B ring edge has the frequency $\Omega_B$ which is very
close to $2:1$ resonance with Mimas frequency $\omega$,
actually $(\Omega_B - \Omega_S)/2\omega = \Omega_B/2\omega -1 \sim 10^{-5} -
10^{-4}$ that corresponds to the accuracy in coordinate position of about $1$ to
$10km$. Thus $\Omega_B$ is located directly in the middle of the
synchronization Arnold tongue where the synchronization effects are the strongest. As a noticeable remark we mention, that it was Christiaan Huygens who discovered both the Saturn's rings (see, e.g. references in book~\citep{fridman}) and the synchronization phenomenon (see, e.g. 
references in book~\citep{pikovsky}).

The paper is composed as follows: the SYNC model of dynamics in the ring is
described in Section 2 (with additional details given in Appendix A, B), 
the numerical and analytical results
are presented in Section 3, the discussion of the results is given 
in Section 4.

\section[]{Description of the SYNC Model of Ring Dynamics}
The SYNC model of the ring dynamics is based on the four main ingredients:
\begin{enumerate}
  \item the individual particle dynamics is given by the Hill equations
        and is considered in a local box as it is proposed by \citet{wisdom},
        the 
	  particles are assumed to be identical, the dynamics is considered only
        in the two-dimensional plane of the ring;
  \item the gravitational force of Mimas is considered as a sequence of periodic kicks
        of fixed amplitude;
  \item the collisions between particles are treated on the basis of the
        mesoscopic multi-particle collision model proposed by Kapral (see e.g 
        \citet{kapral});
  \item the total energy balance (the process where the injection of energy provided by the shear flow \citep{wisdom}
        and the gravitational force of Mimas is equilibrated by dissipation) is
	  ensured via the Nos\`e-Hoover thermostat
	which is broadly used in  molecular dynamics simulations of large ensembles of interacting particles
	(see e.g \citet{hoover,klages,hoover1}).
\end{enumerate}

Let us now describe the elements of the model in more details:

(i) The Hill equations of motion inside a local box \citep{wisdom} are
\begin{equation}
\label{eq1}
    \begin{array}{ll}
\dot{x}=v_x; \;\; \dot{y}=v_y+V_s; \;\;
\dot{v}_x=2\Omega v_y+F_x(t)/m_p; \\
\dot{v}_y=-\Omega v_x/2; \;\;
V_s=-3 \Omega x/2 .
    \end{array}
\end{equation}
where $\Omega=\sqrt{G M_{Saturn}/a_0^3}$ is the Kepler frequency of a particle of mass $m_p$,
$V_s$ is the Kepler shear velocity, $F_x(t)$ is the gravitational force of Mimas along axis $x$
directed to Saturn. Here $v_x, v_y$ are velocities of local motion
in the presence of the shear flow. With $G M_{Saturn} = 3.79 \times 10^{16} m^3/s^2$
and the radius $a_0=1.17 \times 10^8 m$ we have at the edge of B ring
$\Omega=\Omega_S=1.52 \times 10^{-4} s^{-1}$. We normalize all velocities
by a typical value of epicycle velocity $v_{ep}=0.005 m/s$ that gives us 
equations of motion in a 
dimensionless form. After that the distance is measured in 
units of a typical epicycle radius $r_s=v_{ep} /\Omega_S= 32.7 m$
and time $t$ is replaced by $\Omega_S t$.
The local box has the periodic boundary conditions as those used by
\citet{wisdom}.
The size of the box is usually taken as a square $S=5 r_s \times 5 r_s$.
In presence of the Nos\`e-Hoover thermostat (see (iv)) and the shear flow we found convenient
to use the Hamiltonian form of the Hill equations
as it is described by \cite{stewart1}. In this formulation the Hamiltonian of the epicycle motion
has the form $H_{ep}=\Omega I=(v_x^2+v_y^2)/2$ where $I$ is the action of the oscillator motion.
More details are given in Appendix A.

(ii) The computation of the field strength $F_x$ is described in Appendix B.
Because Mimas passes the local box rather fast, the force
in dimensionless units has a form of periodic delta-function
 $f(t)=F_x(t)/(m_p v_{ep}) = \epsilon \sum_l \delta(t-l\tau_M)$
with the period $\tau_M$ defined by the dimensionless orbital 
period of Mimas $\tau_M=2\pi \Omega_S/\omega$
and  dimensionless kick strength $\epsilon=0.64$
corresponding to the fixed choice of the typical epicycle velocity
$v_{ep}=0.005m/s$.
The force in $y$-direction is neglected since it is much 
smaller compared to the force in $x$-direction 
and gives a small change of $v_y$ compared
to the shear velocity.

(iii) The collisions of particles are performed according to the Kapral
algorithm \citep{kapral}. Namely, the whole local box $S$ is divided on
$N_{cel}$ cells. Usually we use about $100 \times 100$ cells with the total number of
particles $N=1000$ corresponding to
$N_{ep}=\pi r_s^2 N/S \approx 125$ particles inside one epicycle circle
and the particle density in a cell being $0.1$. 
After a time $\tau_K$
the relative velocities (with respect to the motion of the center of mass of the cell) of all particles in a given cell are rotated by a random angle. In this way the total momentum and energy inside a
given cell are preserved while the directions of the velocities become mixed. 
It is important to note that during the collision
the velocities inside the cell are taken as 
the total physical velocities of particles,
namely ${\dot x}, {\dot y}$ including the shear velocity.
Thus, due to a finite size of the Kapral cells, the shear velocity
always generates appearance of a spreading 
of local velocities $v_x, v_y$: even if $v_x$ and $v_y$ of two particles coincide before the ``collision'', this is not true for the total velocities ${\dot x}, {\dot y}$, and random rotation of the latter leads to the appearance of non-equal $v_x, v_y$.
In a certain sense the finite size of Kapral cells
physically acts as a finite size of colliding bodies.  
Usually we used $\tau_K=0.5/\Omega_S$ but the
variation of this parameter did not affect the main results.

(iv) In presence of collisions the shear flow and the driving Mimas force
inject additional energy in the system, that is dissipated via different mechanisms, including nonelasiticity of collisions, interactions with dust, etc. In our simplified model, to keep the energy balance,  we use the Nos\`e-Hoover thermostat which is commonly used
for molecular dynamics 
simulations of interacting particles, also in presence of external
fields \citep{hoover,klages,hoover1,chepe}.  In this thermostat,
which mimics a canonical ensemble, 
an additional
friction force acts on a particle $i$ according to
\begin{equation}
\label{eq2} 
\mathbf{\dot p}_i = \mathbf{F}_i - \gamma \mathbf{p}_i \; , \;  \mathbf{\dot
  q}_i = \mathbf{p}_i \; , \,\\
\dot{\gamma} = [\langle \mathbf{p}^2 \rangle/(2m_pT) - 1]/{\tau_H}^2 
\end{equation}
where $\mathbf{p}_i \mathbf{q}_i$  are 
the momentum and coordinate of particle $i$, $\mathbf{F}_i$ is an effective ``friction'' force acting on a
particle due to collisions and external fields,
$\tau_H$ is the relaxation time in the NH thermostat
[usually we used $\Omega_S \tau_H =16$ but we also checked that the variation
of $\tau_H$ does not affect the  synchronization phenomenon
(see examples below)]
and $\langle \mathbf{p}^2 \rangle$ means the average over all $N$ particles,
$T=m_p v_{ep}^2/2$ is a given temperature of the thermostat. 
One can see that the ``friction'' changes the sign with $\gamma$, and the variable $\gamma$ is driven by the deviations of the mean kinetic energy from that at the given temperature $T$, in this way the system is kept near this temperature as it should be for the canonical ensemble.

For the Hamiltonian form of the
Hill equations,  the friction acts only on the action variable
that gives in dimensionless variables 
${\dot I_i} = -\gamma I_i, 
\dot{\gamma}=  [\langle \mathbf{I} \rangle - 1]/{\tau_H}^2$,
where $\langle \mathbf{I} \rangle $ means the averaging over all $N$ 
particles (see Appendix A). The physical origin of the
appearance of such an
effective friction can be attributed to an average friction force acting on 
a relatively large particle as a result of multiple collisions with a dust  of small size particles. 

In a ring with extended size distribution the smaller particles have in
equilibrium generally larger dispersion velocities \citep{salo1992b}.
Since the collisions are inelastic, the system does however not assume a
state of energy equipartition. Typically, the dispersion velocity of the
smallest particles is by a factor of several larger than the one of the
largest particles, depending mildly on the width of the size distribution
and the inelasticity of the particles.

At opposition the perturbing moon induces an equal excess velocity $\Delta
v_x$ to all ring particles, regardless of their size. This means that
compared to collisional equilibrium the large particles have now a higher
excess in random kinetic energy than the small ones. In this sense the
return to equilibrium, mediated by dissipative collisions, affords a
cooling of the large particles relative to the small ones, and thus, an
effective friction on large particles. This friction vanishes in
equilibrium like the Nose-Hoover thermostat. Large particles determine the
dynamical properties of the ring.

Another dissipative process could be an ongoing exchange of ring-matter
 between particles of all size groups due to a balance of coagulation
 and fragmentation (dynamic ephemeral bodies, DEB's,
 \citep{davis1984}) causing an effective dissipation, since the
 composition and destruction of agglomerates are irreversible processes.
%In the rings there are particles of various sizes
%and the Nos\`e-Hoover thermostat models the stabilization forces
%acting on particles of a relatively large  size  $>1 cm$.

It is interesting to note that the epicycle dynamics is
rather similar to motion of charged particles in a magnetic
field \citep{fridman} and due to that there is a certain analogy
with the synchronization of the Larmor motion for two-dimensional electron
gas in magnetic and microwave fields as it was discussed by \citet{chepe}.
The main difference to the present problem is that for the electron gas there is no shear.  

\section[]{Numerical results and their interpretation}

\begin{figure}
% \vspace{302pt}
%\centerline{\epsfxsize=4.2cm\epsffile{fig1a_draft.eps}
%\hfill\epsfxsize=4.2cm\epsffile{fig1b_draft.eps}}
\centerline{\includegraphics[width=4.2cm]{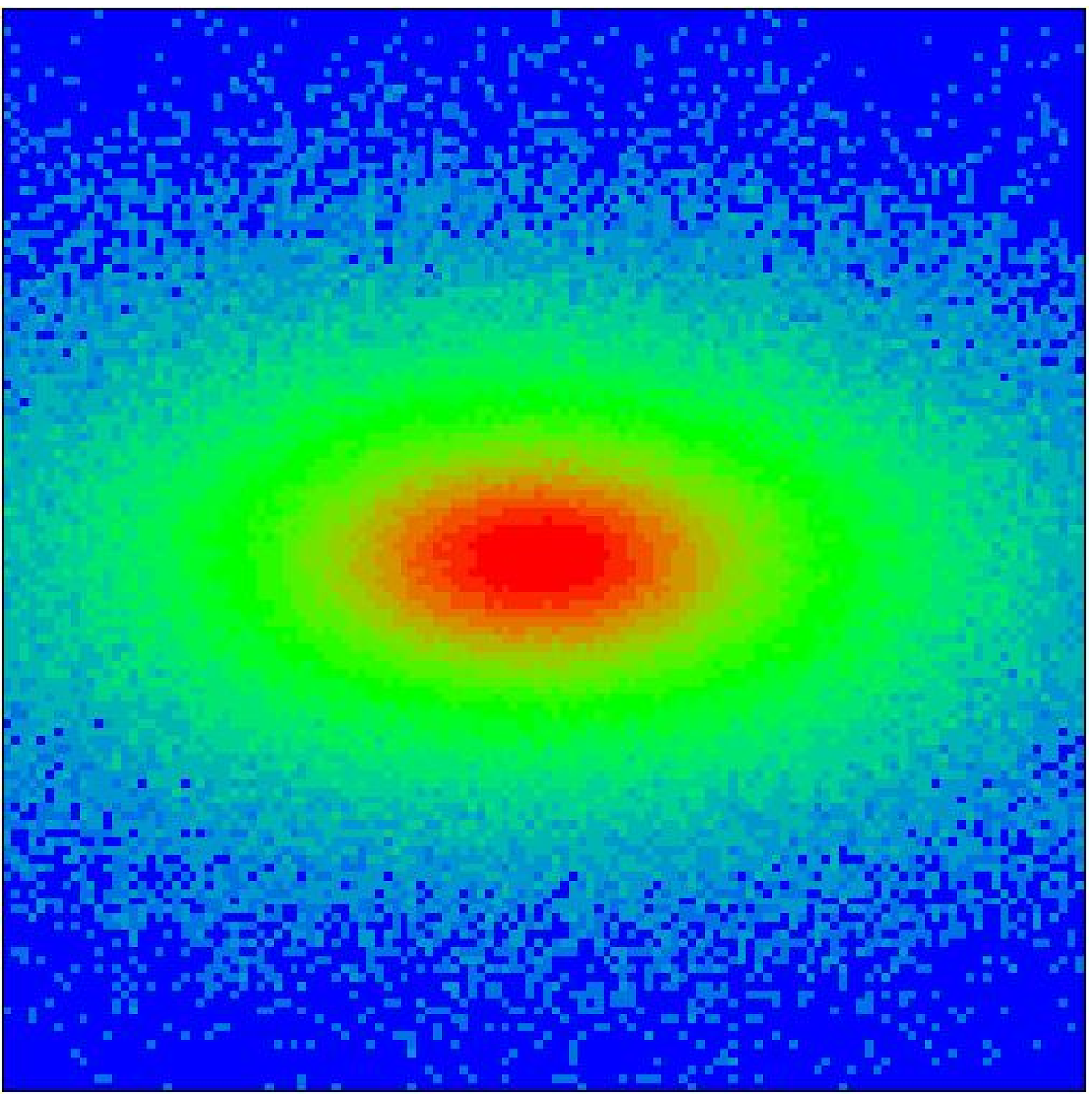}\hfill
\includegraphics[width=4.2cm]{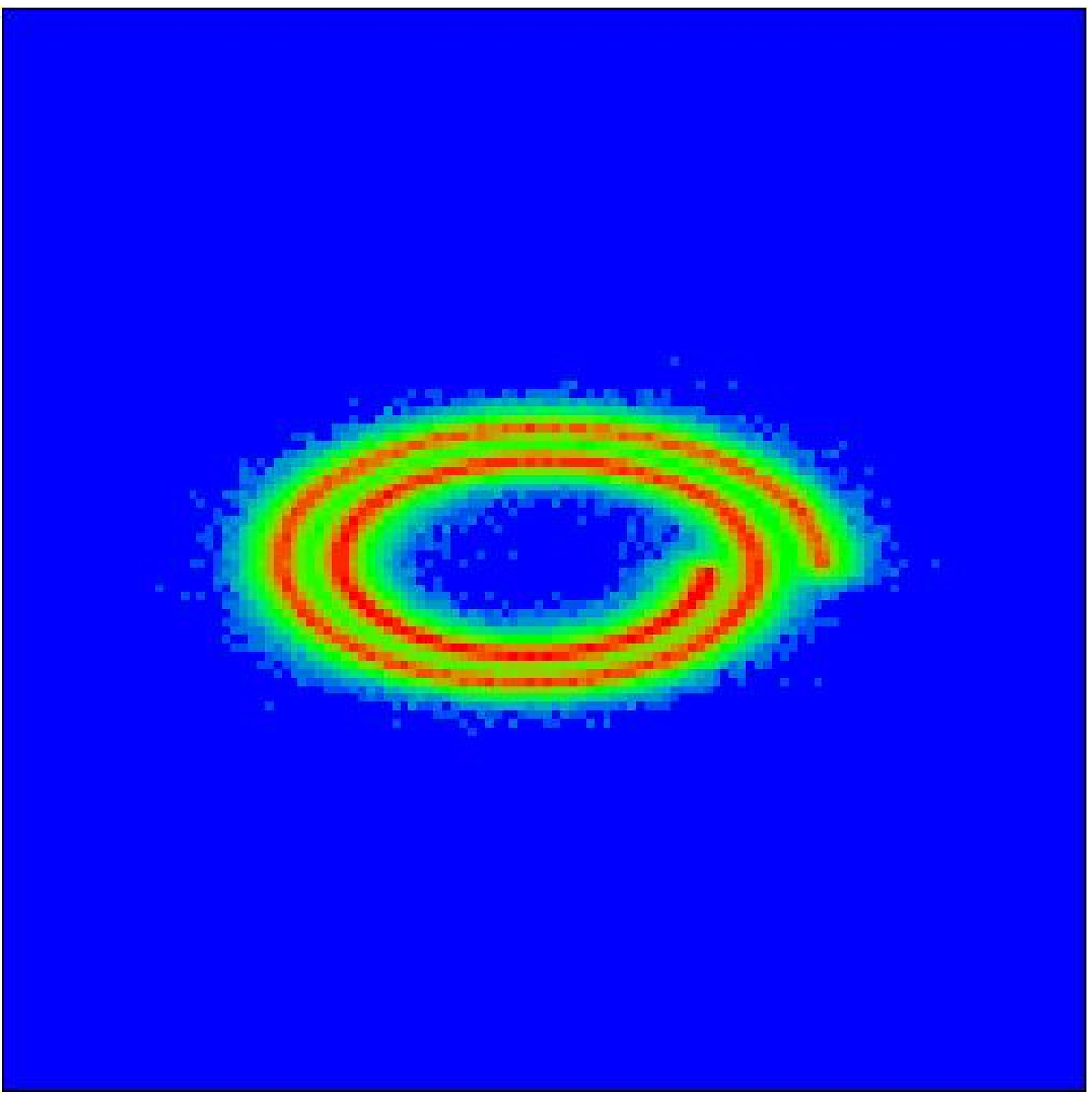}}
 \caption{Density distribution of particles in the ring 
in the plane of local epicycle velocities
$(-3<v_x/v_{ep}<3; -3<v_y/v_{ep}<3)$ obtained by the numerical simulations
with $N=1000$ particles inside the spacial square box
$S=5 r_s \times 5 r_s$ where $r_s$ is the epicycle radius;
the particle density is $\rho=N_p/S =40/r_s^2$.
The rotation frequency ratio is
$\Omega/\Omega_s=\Omega/2\omega=1.15$ (left panel) and 1 (right panel,
synchronized regime);
the dimensionless force amplitude of Mimas is
$\epsilon=0.64$. The number of Kapral cells is
$N_{cel}=100 \times 100 = 10^4$, the Kapral collisions are done
after time $\tau_K=0.5/\Omega_s$; the relaxation time of Nos\`e-Hoover
dynamics is $\tau_H =16/\Omega_s$. The data are averaged over time interval
$0 \leq t \leq 10^4/\Omega_s$. Density is proportional to color
(red/gray for maximum, blue/black for minimum).
 }
\label{fig1}
\end{figure}

The results of numerical simulations for the particle density distribution in
the plane of local epicycle velocities $(v_x,v_y)$ are shown in
Fig.~\ref{fig1} for two values  of $\Omega/2\omega$ ratio between the epicycle frequency
of particles $\Omega$ and the double frequency of Mimas $2\omega$.
For $\Omega/2\omega=1.15$ the distribution of velocities is close to the
Maxwell distribution of elliptical form appearing due to shear and ellipticity
of motion given by the Hill equations (the distribution becomes close
to a symmetric one in rescaled velocities $\tilde{v}_y=2v_y, {\tilde v}_x=v_x$).
The distribution is drastically changed for 
$\Omega/2\omega=\Omega_S/2\omega=1$ :  almost all density is concentrated on a spiral in the
velocity plane. The physical meaning of this phenomenon is the following:
for the resonant ratio $\Omega/2\omega =1$ the phases of the epicycle
rotations become synchronized with the phase of periodic Mimas kicks given by
$f(t)$. While out of the resonance (e.g. $\Omega/2\omega=1.15$) all epicycle
phases are random and independent, in the synchronization regime they are all 
adjusted to the phase of Mimas. As a result,
 Mimas gives a kick in $v_x$, between kicks
the particle velocity decreases doing two rotations over the spiral, then
again it is kicked by Mimas in the same state as at previous kick, and so on. Thus the phases of this rotational motion
are equal to the phase of Mimas and equal to each other. This means that the
particles rotate in synchrony, at each moment of time their rotational velocities are equal both in the amplitude and in the direction.   Therefore, the collisions between 
them effectively \textit{disappear}. There remains only a residual relative velocity, related to
the shear velocity and a
finite size of the Kapral cells (finite size of colliding bodies in the ring).
In the absence of shear flow the synchronization 
can be complete for all particles
as it has been discussed by \citet{chepe}
for the problem of two-dimensional electron gas
in magnetic and microwave fields. 

\begin{figure}
%\centerline{\epsfxsize=8.4cm\epsffile{fig2_draft.eps}}
\centerline{\includegraphics[width=8.4cm]{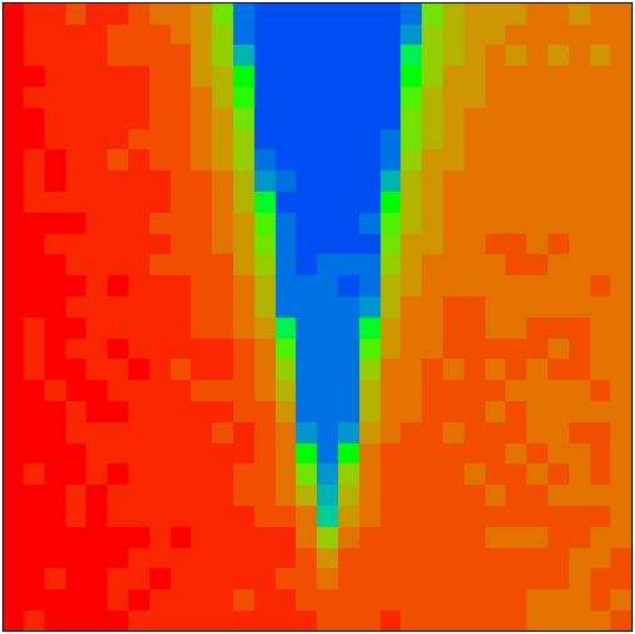}}
  \caption{Dependence of the rescaled diffusion rate
  $D/{\tilde D}_0$ shown by color on 
  the rescaled frequency $\Omega/2\omega$ (horizontal axis)
   and driving force strength $\epsilon$ (vertical axis)
  for the range $0.85 \leq \Omega/2\omega \leq 1.15$
  and $0 \leq \epsilon \leq 0.7$.
  The color is proportional to
  $D/{\tilde D}_0$ with red/gray for maximum value
  ($D/{\tilde D}_0=1.26$) and blue/black for minimum
  ($D/{\tilde D}_0=4 \times 10^{-4}$), here
  ${\tilde D}_0$ is the diffusion rate
  at $\epsilon=0, \Omega=\Omega_s$
  (also $D/D_0=1.7 \times  10^{-5}$ with $D_0=r_s^3 \Omega_s$).
  Data are obtained for time $t \leq 10^4/\Omega_s$, $N=1000$,
  $N_{cel}=200 \times 200$, other parameters are as in Fig.1.
  }
\label{fig2}
\end{figure}

Since the collisions are significantly reduced for the synchronization
regime, the diffusion rate $D=\langle x^2\rangle/t$ is also reduced by orders of magnitude
compared to its typical value $D_0=r_s^2 \Omega_S$. This is clearly seen in
Fig.~\ref{fig2} which gives the dependence of $D$ on $\Omega$ and $\epsilon$.
The diffusion suppression takes place inside the Arnold tongue where
the epicycle rotation is synchronized with the Mimas phase. According to the
data of Fig.~\ref{fig2} the synchronization takes place inside the frequency 
range
\begin{equation}
\label{eq3}
   |\Omega/2\omega - 1| \leq s \epsilon
\end{equation}
with the numerical value of the constant $s \approx 0.08$.
According to the synchronization theory \citep{pikovsky} the synchronization
region is given by the dimensionless amplitude of the driving force
which is $2 \epsilon/(4\pi \text{max}(v_x))$ that gives $s \approx 0.08$ 
since we have $\text{max}(v_x) \approx 2$ (see Fig.~\ref{fig1}).
Thus the numerical dependence is in agreement with the analytical
synchronization theory. The variation of $\Omega$ is related to the position
of particle inside the ring. For example, $\Omega/2\omega=1.05$
corresponds to the distance $\Delta x \approx 3a_0(\Omega/2\omega -1)/2
\approx 9000 km$ from the outer B edge in direction to Saturn.
It is interesting to note that this is of the order of 
the size of Cassini division.

\begin{figure}
\centerline{\includegraphics[width=5.cm]{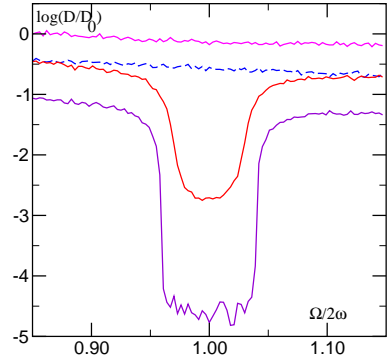}}
  \caption{Dependence of the rescaled diffusion rate
  $D/D_0$ on 
  the rescaled frequency $\Omega/2\omega$. The full curves are for
  $\epsilon=0.64$ with $N_{cel}=50 \times 50, 100 \times 100, 200 \times 200$
  from top to bottom; the
  dashed curve is for $\epsilon=0$ and $N_{cel}= 100 \times 100$.
  Data are obtained for time $t \leq 10^4/\Omega_s$, $N=1000$.
  Logarithms are decimal.
\label{fig3}
  }
\end{figure}

An interesting property of the relation (\ref{eq2}) is its independence of the
relaxation time scale $\tau_H$. Physically, this means that $\tau_H$ only
determines the time scale on which the synchronization is reached 
but it does not  affect the domain of synchronization. This is in agreement with our
numerical checks which show that at the Mimas value of $\epsilon=0.64$
the synchronization window shows less than $10\%$ variation when
$(\Omega_S \tau_H)^2$ varies from $0.1$ to $10^{-5}$. For $(\Omega_S
\tau_H)^2=1$ this window is increased by about 30\%.

\begin{figure}
\centerline{\includegraphics[width=5cm]{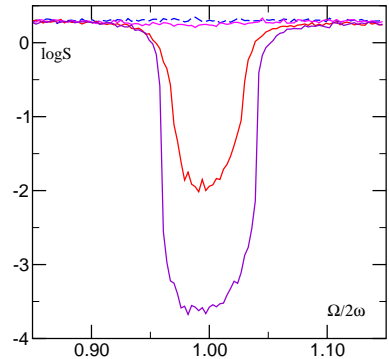}}
  \caption{Dependence of the synchronization parameter
  $S$ on the rescaled frequency $\Omega/2\omega$. 
  Parameters are the same as in Fig.3.
  Logarithms are decimal.
\label{fig4}
  }
\end{figure}

The variation of the diffusion rate $D(\Omega)$ when the number of Kapral
cells $N_{cel}$ is changed is shown in Fig.~\ref{fig3}. For
$N_{cel}=50\times50$ the collisions happen rather often and there are no signs
of synchronization. For $N_{cel}=100\times100$ the synchronization sets in and
the diffusion drops inside the synchronization window.
A further increase up to $N_{cel}=200\times200$ gives much stronger drop
of the diffusion inside the synchronization window while its size is only slightly
increased. Outside of this window the diffusion scales as
$D \propto 1/N_{cel}$. This is rather natural since $D$ is proportional to
the density of particles inside the cell so that 
$D \sim r_s^2 \Gamma \sim r_s^2 N/(N_{cel} \tau_K)$ where $\Gamma$ is the
effective collision rate. We note that in the non-synchronized regime
the diffusion rate per orbital period is rather large being 
$2\pi D/(r_s^2\Omega_S) \approx 3$ (at
$N_{cel}=100 \times100$)  and 1 (at $N_{cel}=200 \times 200$). These values
approximately correspond to the typical conditions for particles inside B ring.

Another signature of synchronization can be expressed via $\;\;\;\;\;\;$ the synchronization
parameter $\;\;\;\;\;\;$ $S= \sum_{i<j} (\mathbf{v}_i - \mathbf{v}_j)^2/(N^2 v_{ep}^2/2)$.
Its dependence on frequency is shown in Fig.~\ref{fig4}.
Inside the synchronization regime $S$ drops by almost 4 orders of magnitude.
This means that due to synchronization the relative collision velocities of particles are 
very small and therefore the diffusion is also very small. At the same time
the velocity difference remains finite due to finite size of the cells
(collision bodies) and the shear flow. In absence of the shear flow 
the collisions disappear completely (see also \citep{chepe}).

For the chosen value $v_{ep}=0.005m/s$ we have $\epsilon=0.64$ 
(see Appendix B) that according to (\ref{eq2}) and the data
of Figs.~\ref{fig2},\ref{fig3},\ref{fig4} give the synchronization border
$\Omega/2\omega \approx 1.05$. This corresponds approximately
to $x_S=9000km$ distance from the exact resonance 2:1. The observations give this
distance to be about $x_B \approx 1 - 10 km$ that is significantly smaller than the
synchronization border. Of course, the value of $v_{ep}$ is not known exactly
and may be a factor $2$ to $10$ different from the value chosen above.
But even if the actual value is  10 times larger than still
$x_S$ remains by 2 orders of magnitude larger than the observed B edge
position $x_B$. 

\begin{figure}
\psfrag{xlabel}[c][c]{position $x/r_S$}
\psfrag{ylabel}[c][c]{time $\Omega_s t$}
%\centerline{\epsfxsize=8.4cm\epsffile{fig5_draft.eps}}
\centerline{\includegraphics[width=8.4cm]{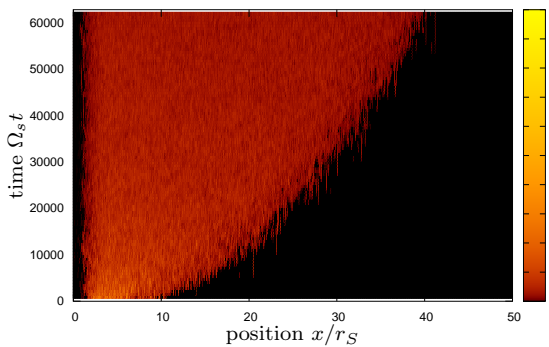}}
  \caption{Dependence of density of particles (in arbitrary units) on time 
  $\Omega_St$ (vertical  axis) and position in the ring
  $x/r_S$ (horizontal axis) for $\epsilon=0.6$, $\Omega/2\omega=1.1$
  and zero frequency gradient $g=0$..
  There are $N_{cel}=1200 \times 120$ in the whole space box
  $S=50 r_S \times 5 r_S$; $\tau_K=0.5/\Omega_s$,
  $\tau_H=4.5/\Omega_S$.
  Initially there are $N=60$ particles in the left box $S=5 r_S \times 5 r_S$
  and this number is kept constant during the computations
  till the finite moment of time $\Omega_St =6.28 \times 10^4$
  when there are 305  particles in total.
\label{fig5}
  }
\end{figure}

We explain this disagreement in the following way. Taking the value
$\epsilon=0.64$ we assume that  the particles appear initially in
the non-synchronized part of the ring, let say  with 
$\Omega/2\omega \approx 1.2$. Due to collisions these particles diffuse
closer and closer to the synchronization border at 
$\Omega/2\omega \approx 1.05$. Behind this border the diffusion rate drops 
$D(\Omega)$ drastically due to the synchronization phenomenon described above.
This sharp drop of $D$ creates a diffusive shock wave which 
continues to propagate slowly inside the synchronization region
since there the diffusion remains finite due to the shear flow 
(see Figs.~\ref{fig2},\ref{fig3}). The edge size of this diffusive shock wave
should be of the size of few epicycle radius $r_s$ since as soon as the
distance between particles becomes larger than $r_s$ the collisions between
them completely disappear due to a frozen nature of epicycle motion
(in a close similarities with particles in a magnetic field, see
e.g. \citep{fridman}). In this way the gradient of density slowly moves inside
the synchronization domain keeping the sharpness of the edge of a few
epicycle radius $r_s$. In the synchronization domain the diffusion is minimal
at the exact resonance $\Omega/2\omega=1$ since the synchronization effect
is the most strong there (see Figs.~\ref{fig2},\ref{fig3},\ref{fig4}).
Therefore, the edge of synchronized particles spends the most of the time
at that place with the minimum of diffusion. This is actually the place 
where the outer B ring edge is observed now. 

\begin{figure}
\psfrag{xlabel}[c][c]{position $x/r_S$}
\psfrag{ylabel}[c][c]{time $\Omega_s t$}
%\centerline{\includegraphics[width=8.4cm]{fig6_draft.eps}}
\centerline{\includegraphics[width=8.4cm]{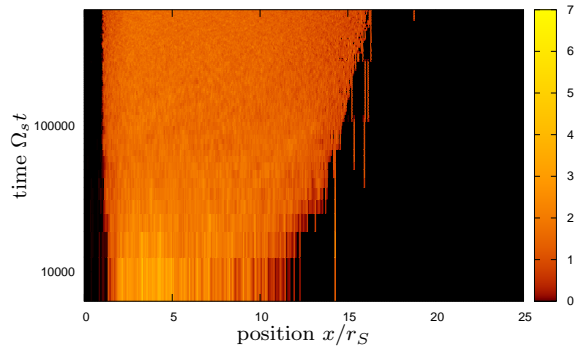}}
  \caption{Dependence of density of particles (in arbitrary units) on time 
  $\Omega_St$ (vertical  axis in logarithmic scale) and position in the ring
  $x/r_S$ (horizontal axis) for $\epsilon=0.6$, the frequency gradient in
  space is $g=0.002$ with $\Omega/2\omega=1.05$ at $x=0$ and
  $\Omega/2\omega=1$ at $x/r_S=25$ corresponding to the outer B ring edge.
  There are $N_{cel}=1200 \times 120$ in the whole space box
  $S=50 r_S \times 5 r_S$ (only half is shown); $\tau_K=0.5/\Omega_s$,
  $\tau_H=4.5/\Omega_S$. 
  Initially there are $N=60$ particles in the left box $S=5 r_S \times 5 r_S$
  and this number is kept constant during the computations
  till the finite moment of time $\Omega_St =6.28 \times 10^5$
  when there are 305 particles in total.
\label{fig6}
  }
\end{figure}

To give more justification to the above picture we performed 
extensive numerical simulations of front propagation of particles 
in the SYNC model. With this aim we introduced a gradient
of the frequency $\Omega$ with the distance $x$
so that $\Omega(x)=\Omega_0-gx$ where $\Omega_0 \approx 1.1$ 
is some initial value and $g \sim 0.002$ is the frequency gradient
per unit of epicycle radius in dimensionless units.  
Examples of the numerical simulations
are shown in Figs.~5,6. In the absence of the gradient, in the non-synchronized 
regime there is a diffusive spreading along $x$ as it is clearly seen 
in Fig.~\ref{fig5}. The typical diffusive profile $x\propto \sqrt{t}$ is
clearly seen. The computations are done at the fixed constant particle density
$\rho$ in the left space box $5 r_S \times 5 r_S$ of the total longitudinal
space box $S=50 r_S \times 5 r_S$ with the number of Kapral cells
$N_{cel}=1200 \times 120$. This is reached by adding new particles inside the
left space box during the diffusive spreading. At such a density $\rho r_S^2
=60/25$ and such a value $N_{cel}$, the local diffusion rate
at $\Omega/2\omega \approx 1.1$ is $D/D_0 \approx 0.01$
and during the time $\Omega_S t =6.28 \times 10^4$
the diffusion propagates on a distance $x \approx \sqrt{D t} \approx 25 r_S$
that is in a good agreement with the data of Fig.~\ref{fig5}.  
  
\begin{figure}
\psfrag{xlabel}[c][c]{position $x/r_S$}
\psfrag{ylabel}[c][c]{density}
%\centerline{\epsfxsize=8.4cm\epsffile{fig7.eps}}
\centerline{\includegraphics[width=8.4cm]{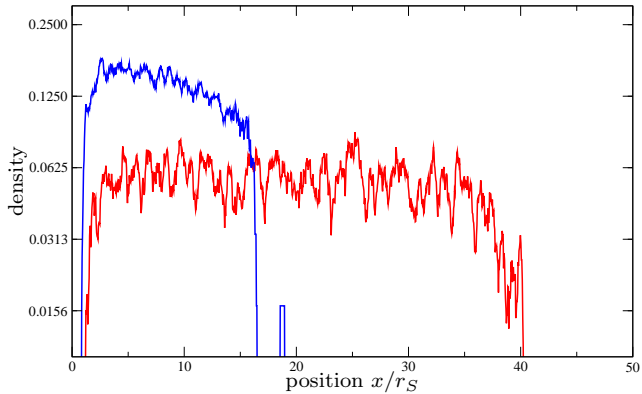}}
  \caption{Dependence of  density of particles 
  (arbitrary units in logarithmic scale) 
  on position in the ring $x/r_S$ for $\epsilon=0.6$ at the final time
  $\Omega_s t=6.28 \times 10^4$ (red/gray for parameters of Fig.~\ref{fig5})
  and $\Omega_s t=6.28 \times 10^5$  (blue/black for parameters of Fig.~\ref{fig6}).
\label{fig7}
  }
\end{figure}

The situation is drastically different in presence of the frequency gradient
$g=0.002$ when the particles enter inside the synchronization window
$\Omega/2\omega -1 < 0.05$ as it is shown in Fig.~\ref{fig6}. Even if the
total computation time here is 10 times larger than in Fig.~\ref{fig5}
the front propagation becomes very slow around $x/r_S \approx 17$ 
since diffusion drops strongly inside the synchronization window
going down to very small by finite value 
$D/D_0 \approx 3 \times 10^{-6}$ inside the left space box 
$5r_S \times 5r_S$ with $60$ particles. 

Of course, the value of the gradient chosen here is much larger than its real
value $g \sim 3 r_S/2 x_s \sim 15m /1.17 \times 10^5 km \sim 10^{-7}$.
Such small values of the gradient are not accessible for nowadays
computer simulations.
However, a more smooth, adiabatic variation of the orbital frequency $\Omega$
on a scale of an epicycle radius should make the picture of the diffusive
shock wave moving inside the synchronization domain to be even 
better justified.

The density $\rho(x)$ dependence on $x$ at a final moment of time 
(averaged over 1\% of total time) is shown in Fig.~\ref{fig7}.
for the cases of Figs.~\ref{fig5},\ref{fig6}.
In the case of synchronization (Fig.~\ref{fig6}) there is a sharp drop of
density on a scale of 2 epicycle radius $r_S$. In the non-synchronized case
(Fig.~\ref{fig5}) the decrease of density goes in a more smooth way
(on a scale of about $10 r_S$). The difference of scales in two cases
is not so large since in both cases the diffusion is zero in the region
without particles. However, the most important difference is that
the rapid diffusive propagation goes unlimitedly in the non-synchronized
case while inside the synchronization window the propagation
front moves very slowly with the formation of sharp density drop
on a scale of about $2r_S$. The front stays the longest time at the place
where the diffusion is minimal that corresponds to the center
of the synchronization window at $\Omega=2\omega$.
For the value of $r_S \propto 1/v_{ep} \approx 36m$
given above this gives the size of the edge 
$\Delta x_e \approx 2 r_S \approx 70 m$. This value is in a good agreement
with the observation data which give $\Delta x_e \approx 10m$ especially
if we take into account that the average value $v_{ep}$ is known only by an
order of magnitude.

\section{Discussion}

Our studies based on the SYNC model of dynamics in the rings of Saturn
show the emergence of synchronization in the vicinity of the outer B ring
edge. Like the Maxwell demon this synchronization makes the 
epicycle motion of particles to be synchronous that practically
eliminates collisions between them. This gives a suppression of diffusion
by orders of magnitude and a formation of a diffusive shock wave
slowly propagating inside the synchronization domain.
The size of this front or the edge of the ring is of the order of a few 
epicycle radius being of about of a few tens of meters.
This is in agreement with the present observation data which give its size
to be about 10 meters. The size of the synchronization domain
created by Mimas is of the order of $4000km$ (remarkably, the total number of synchronized
particles, which can be estimated as $>10^{12}$, is huge).
The front moves most slowly in the center 
of the synchronization domain so that it is most probable to
observe it at the exact resonance 2:1 between the edge frequency and the frequency
of Mimas. The observations show that the actual position is close to this
value with a relative accuracy of $10^{-4} -10^{-5}$.

Similar synchronization effects should exist at other resonances
with other satellites. Our preliminary data give a similar picture for the
outer A ring edge which is closely located to the 7:6 resonance
with Janus (here the dimensionless kick strength is $\epsilon \approx 1.19$
but the relative frequency size of the synchronization window is approximately
twice smaller due to higher order of the resonance). 

The synchronization mechanism proposed here can also be responsible for
existence of very narrow planetary rings. Indeed, if initially particles 
are distributed inside the resonance then those which are inside the
synchronization window will remain there practically forever since 
the collision induced diffusion is switched off, while those outside
of the window will diffuse away living a narrow ring of particles inside the
synchronization window.

The SYNC model used in this paper is based on several significant simplifications of the real particle dynamics. The use of the mesoscopic Kapral method for incorporating the collisions appears to imply no additional physical mechanisms, but the basic physics of the collisions remains hidden inside the parameters of the Kapral method - the number of the cells and the frequency of the reshuffling of velocities. To make the calculations more realistic, one needs to incorporate realistic collision models with reasonable mechanical properties of the particles, to include a distribution of the sizes, etc. Such an extension goes far beyond the scope of our preliminary study. Thus we can hardly make direct predictions, in particular, compare properties of A and B ring edges. 

Moreover, it appears that another simplification of our computational model --  the use of the Nos\`e-Hoover thermostat -- is less ``repairable'' and requires a further justification. Indeed, if there is some average dissipation due to particle collisions, this justifies the use of the thermostat for modeling the saturation of the energy pumped to the system due to shear. On the other hand, in our setup in the synchronized state collisions are rare so at the first glance there is no mechanism for energy saturation. Here it is important to mention, that we
restricted our analysis to an ensemble of identical particles. For a particular situation in the outer B ring this means that we consider only large particles that have low characteristic epicyclic velocities. If there is a whole range of particles of different sizes, they are all subject to resonant kicks by Mimas, but because their characteristic epicyclic velocities are different, the effect of the kicks is also different. If we assume rough equipartition of epicylic kinetic energies, then smaller particles have larger velocities, therefore for them the effective forcing parameter $\epsilon$ is smaller, and as a result they only weakly synchronize or not synchronize (if they lie near the bottom of Fig.~\ref{fig2}). Collisions with these randomly moving particles may provide
an additional dissipation that justifies the use of  the
Nos\`e-Hoover thermostat. 
Further investigations
(which however go far beyond the scope of this paper)
of different dissipation effects influencing the energy balance are needed to clarify this issue. 

Another simplification made -- a consideration of a relatively small box of particles -- is also crucial. 
Indeed, parts of the ring at different angular coordinates are statistically equivalent, but the
dynamical equivalence is broken by the influence of Mimas, as the latter kicks the particles at different phases. Thus the synchrony of the velocities may be only local, i.e. the velocities of particles are synchronized with the local phase of the Mimas, but this phase changes gradually along the ring angular coordinate. Because of the differential rotation, the particles synchronized at different phases will mix, but this effect is not taken into account in the box model, it has to be addressed in future studies.

In this respect the important issue is that of how the synchronization of particles' velocities could be tested experimentally. Indeed, the theory above predicts a drastic narrowing of the width of the velocities distribution, cf. Fig.~\ref{fig1}. However, because the velocities themselves are very small, there is no much hope to
observe the distribution of them directly, e.g. by Doppler measurements or image analysis. Thus one has to rely on implicit observations. For example, one may expect that the adjustment of velocities influences other dynamical features recently observed in the rings, like propeller structures (\cite{schmidt,spahn}). However, for such an implicit test one has to find such structures quite near to the sharp edges described above. 
Remarkably, near the outer edge of ring B a scrambled pattern is observed in recent Cassini images
( see URL: http://saturn.jpl.nasa.gov/photos/imagedetails/\ index.cfm?imageId=2984 ) that is probably due to a gravitational clumping of 
particle (cf. simulations by \citet{lewis05} of a similar structure at the Encke gap). We can speculate that the synchronization of large particles would generally enhance their ability to clump due to gravitational and adhesive forces, because their relative collision velocities become rather small.  On the other hand, if smaller, non-synchronized particles, leave the ring and enter the gap, where large particles are missing, their diffusion will be reduced as well, because the collisions between small particles are rare. Observation of such a gradient in the size distribution of particles near the edge may also be interpreted as an implicit confirmation of the suggested mechanism.

We note that in the literature several mechanisms for explanation of sharp edges of rings have been discussed. 
One process to form and maintain a sharpness is a shear reversal induced by a strong gravitational perturbation near a resonance with a moon. In the unperturbed ring Keplerian shear leads to an outward viscous transport of angular momentum. The distortions induced by a perturber may locally reverse the shear and, when integrating over the azimuth, to a net angular momentum transport that is directed radially inward 
\citep{bgt82,bgt83,bgt89,mosqueira}. 
This process can lead to an inward migration of material from the close vicinity of the resonance location with the moon, and in this way clear a gap at this radial position in the ring.
 Recently Lewis et al. reported on a ``negative diffusion'' mechanism, where the particles migrate to areas of high density (\cite{leezer}). In their calculations, however, only a single pass by the moon was simulated so no resonant effects, like those we focus on in this study, were possible.
Probably, further observations, together with further extensive numerical simulations, will allow one to test which mechanism is responsible for the sharp edges of the rings.  It appears highly desirable to compare various setups in numerical simulations as well.

Finally, we note that, as discussed by \citet{chepe},  
the elimination of collisional diffusion
may appear also for charged particles in a magnetic field
like 2D electron gas (see \citet{mani2002,zudov2003})
and electron and ion clouds in a Paul trap (see, e.g., \citet{drewsen}).
Due to that it can be rather interesting to try to make
laboratory experiments with traps (see e.g. \citet{traps})
which would allow to model rings of Saturn in laboratory
experiments with cold ions.

\section*{Acknowledgments}

This  research is supported in part
by the projects MICONANO and NANOTERRA of the  ANR France (for DLS),
also DLS thanks Univ. of Potsdam for hospitality 
during the final period of this work.

\appendix

\section[]{The Hamiltonian form of the Hill equations}

Here we describe the Hamiltonian form of the Hill equations. According to \cite{stewart1}, the Hill equations (\ref{eq1}) can be viewed as a Hamiltonian system with
\begin{equation}
H=\frac{1}{2}[(p_x+\Omega y)^2+(p_y-\Omega x)^2]-\frac{3}{2}\Omega^2x^2- x F_x(t)/m_p
\label{eqA1}
\end{equation}
For the simulation and the analysis another representation, also given by \cite{stewart1}, where the epicyclic and shear motion are effectively separated, is more convenient. One introduces canonical variables $I,\phi,P,Q$ according to
\begin{align}
x&=\frac{2}{\Omega}P-\sqrt{\frac{2I}{\Omega}}\cos \phi,\qquad
y=Q+2\sqrt{\frac{2I}{\Omega}}\sin \phi\label{eqA21}\\
\dot x&=\Omega \sqrt{\frac{2I}{\Omega}}\sin \phi,\qquad
\dot y=-3P+2\Omega \sqrt{\frac{2I}{\Omega}}\cos \phi\label{eqA22}
\end{align}
In these variables the Hamilton function reads
\begin{equation}
H=\Omega I-\frac{3}{2}P^2-
\left(\frac{2}{\Omega}P-\sqrt{\frac{2I}{\Omega}}\cos \phi\right)F_x(t)/m_p
\label{eqA3}
\end{equation} 
Canonically conjugated variables $I,\phi$ are the action-angle variables for the epicyclic motion. When introducing the Nos\`e-Hoover thermostat, we adjust variable $I$ only, modelling in this way the balance of this part of the total energy. Canonically conjugated variables $P,Q$ describe the shear, the conserved quantity $P$ corresponds to the conservation of the angular momentum.
Note that according to  (\ref{eqA21}), variables $P$ and $Q$ can be viewed as ``center of mass'' coordinates for the rotating particle, it is especially convenient to calculate the diffusion rate in $x$-direction in terms of the diffusion constant for $P$, because in the absence of collisions this quantity is exactly conserved.

\section[]{Derivation of Mimas's kick force}

Here we derive the strength of kick force produced by Mimas on the epicycle motion
of particles inside the Saturn ring B. Consider the effect of a gravitational action of the Mimas having mass $m_m$ and semi-major axis $a_m$ on a particle having axis $a$ (both orbits are nearly circular). In a frame fixed with the particle, the graviational acceleration in the outer direction is $\dot v_x=GM_m d^{-2}\cos\psi$ where $d$ and $\psi$ is the distance and the angle from the particle to the Mimas. Denoting the angle from Saturn to Mimas as $\phi=(\Omega-\Omega_m)t$ where $\Omega_m$,$\Omega$ are Kepler frequencies, we
can write $\cos\psi=d^{-1}(a_m\cos\phi-a)$ and $d^2=(a_m\sin\phi)^2+(a_m\cos\phi-a)^2$.    The total change of the velocity component $v_x$ due to the force by Mimas is given by the integral $\int dot v_x$. We attribute this change to a ``kick'', because the force is non-zero only when Mimas is close to the particle. The resulting expression is
\begin{equation}
\Delta v_x =\frac{G m_m}{(\Omega-\Omega_m)a^2} \int_{-\pi}^{\pi} \frac{r^2(\cos\phi-r)\;d\phi}{(1+r^2-2r\cos\phi)^{3/2}}
\label{eq:B1}
\end{equation}
This calculation gives $\Delta v_x\approx 3.2\cdot 10^{-3}m/s$. When 
normalizing by a typical epicycle velocity $5\cdot 10^{-3}m/s$ we get a numerical value of our parameter $\epsilon\approx 0.64$.

\bsp

\label{lastpage}

\end{document}